\documentclass[12pt]{article}
\usepackage{titlesec}
\usepackage{epsfig}
\usepackage{graphicx}
\usepackage{graphics}
\usepackage{amssymb}
\usepackage{amsmath}
\usepackage{amsfonts}
\usepackage{color}
\usepackage{textcomp}
\usepackage{latexsym}
\usepackage{cite}
\usepackage{titletoc}
\usepackage{tikz}
\usepackage{xcolor}
\usetikzlibrary{3d}
\usepackage{csquotes}
\usepackage{caption}
\usetikzlibrary{calc} 
\usetikzlibrary{angles,quotes}
\usepackage{csquotes}
\titleformat{\section}{\normalfont\Large\bfseries}{\thesection.}{1em}{}
\titlecontents{section}[0em]{\bfseries}{\thecontentslabel. }{}{\titlerule*[1pc]{.}\contentspage}
\usepackage[margin=2.5cm,footskip=0.6cm]{geometry}
\begin{document}

\begin{center}
{\Large{\bf Virial Equation of State for a Granular System}} \\
\ \\
\ \\
by \\
Subhanker Howlader and Prasenjit Das\footnote{prasenjit.das@iisermohali.ac.in}\\
\textit{Department of Physical Sciences, Indian Institute of Science Education and Research Mohali, Knowledge City, Sector 81, SAS Nagar, Punjab 140306, INDIA}
\end{center}

\begin{abstract}
\noindent The equation of state for an ideal gas is simple, which is $P=nk_{\rm B}T$. In the case of imperfect gases where mutual interactions among the constituents are important, pressure $P$ can be expressed as the series expansion of density $n$ with appropriate coefficients, known as virial coefficients $B_m$. In this paper, we have obtained the first four virial coefficients for a model interaction potential $\Phi(r)$ using multidimensional Monte-Carlo integration and importance sampling methods. Next, we perform molecular dynamics simulations with the same $\Phi(r)$ for a many-particle system to obtain $P$ as a function of $T$ and $n$. We compare our numerical data with the virial equation of state.
\end{abstract}

\newpage
\section{Introduction}\label{sec1}
Understanding the equilibrium properties of gases has been a centuries-old pursuit, tracing its origins from Boyle's (1662) to Charles's (1787) laws~\cite{bb09}. The combination of these with Gay-Lussac's and Avaogadro's laws led to the synthesis of the ideal gas law: $PV=Nk_{\rm B}T$ by Clapeyron in 1834. The Maxwell's speed distribution and finite size correction of \textit{van der Waals} at high pressure improve our understanding further. For ideal gases, there is no mutual interaction among the gas particles. However, in reality, the behaviors of gases are more intricate due to mutual interactions among the particles. Therefore, we need to incorporate the contribution from the interaction potential $\Phi(r)$ into the equation of state~(EoS). Historically, this has been done through the \textit{cluster expansion} method by Mayer and his collaborators~($1938-1951$), known as \textit{Mayer cluster expansion}~\cite{pb21}. They obtained the EoS as a series of the form~\cite{dm76}
\begin{equation}
\label{Eq1}
\frac{P}{k_{\rm B}T}= B_{1}(T)n + B_{2}(T)n^{2} + B_{3}(T)n^{3} + B_{4}(T)n^{4} + \cdots,
\end{equation}\\
where $n(=N/V)$ is the density. The eq~(\ref{Eq1}) is known as the virial equation of state, $B_{m}(T)$ is the temperature~($T$) dependent $m^{\rm th}$ virial coefficient, and $k_{\rm B}$ is the Boltzmann constant.

The virial coefficient $B_{m}(T)$ involves the interaction potential $\Phi(r)$ through the Mayer function $f_{ij}\equiv f(r_{ij})=\exp\{-\beta \Phi(r_{ij})\} - 1$, where $\beta=(k_{\rm B}T)^{-1}$ and $r_{ij}$ is the distance between the centers of a pair of particles $i$ and $j$. The mathematical expressions for the first four virial coefficients are as follows~\cite{pb21,dm76,ss10}:
\begin{align}
\label{Eq2}
B_1(T)&=\frac{1}{V}\int d^3 r_1=1~
\begin{tikzpicture}
  [scale=.6, baseline={(0,0)}, yshift=0.2cm]
  % Draw the bead
  \filldraw (0, 0) circle (0.15);
\end{tikzpicture}\\
\label{Eq3}
B_2(T)&=-\frac{1}{2 V} \iint f_{12} d^3 r_1 d^3 r_2=-\frac{1}{2V}~
\begin{tikzpicture}[scale=.6, baseline={(0,0)}, yshift=0.2cm]
    % Draw the beads
    \foreach \x in {0, 2}
        \foreach \y in {0}
            \filldraw (\x, \y) circle (0.15);
    % Draw the connecting arm
    \draw (0, 0)--(2, 0);
\end{tikzpicture}\\
\label{Eq4}
B_3(T)&=-\frac{1}{3 V} \iiint f_{12} f_{13} f_{23} d^3 r_1 d^3 r_2 d^3 r_3=-\frac{1}{3 V}
\begin{tikzpicture}[scale=.55, baseline={(0,0)}, yshift=-0.7cm]
    % Define the coordinates of the vertices
    \coordinate (A) at (0, 0);
    \coordinate (B) at (2, 0);
    \coordinate (C) at (1, 2);
    % Draw the triangle edges
    \draw (A) -- (B) -- (C) -- cycle;
    % Draw beads at each vertex
    \foreach \coord in {A, B, C}
        \foreach \i in {0,120,240}
            \filldraw (\coord) circle (0.16);
\end{tikzpicture}\\
\label{Eq5}
B_4(T)&=-\frac{1}{8V}\iiiint \left[3f_{13}f_{14}f_{23}f_{24} + 6f_{12}f_{13}f_{14}f_{23}f_{24}+f_{12}f_{13}f_{14}f_{23}f_{24}f_{34}\right]d^3r_{1}d^3r_{2}d^3r_{3}d^3r_{4}\nonumber \\ &= -\frac{1}{8V}\left(3\begin{tikzpicture}[baseline={(0.5,0.5)}, yshift=0.7ex]
        % Draw the first small square with circular beads at each corner
        \draw (0, 0) rectangle (1, 1);
        \foreach \x/\y in {0/0, 0/1, 1/0, 1/1}
            \filldraw (\x, \y) circle (0.1);
    \end{tikzpicture}
    + 6\begin{tikzpicture}[baseline={(0.5,0.5)}, yshift=0.7ex]
        % Draw the second small square with circular beads and diagonal line
        \draw (0, 0) rectangle (1, 1);
        \foreach \x/\y in {0/0, 0/1, 1/0, 1/1}
            \filldraw (\x, \y) circle (0.1);
        \draw (0, 0) -- (1, 1);
    \end{tikzpicture}
    + \begin{tikzpicture}[baseline={(0.5,0.5)}, yshift=0.7ex]
        % Draw the third small square with circular beads and two diagonal lines
        \draw (0, 0) rectangle (1, 1);
        \foreach \x/\y in {0/0, 0/1, 1/0, 1/1}
            \filldraw (\x, \y) circle (0.1);
        \draw (0, 0) -- (1, 1);
        \draw (0, 1) -- (1, 0);
    \end{tikzpicture}
    \right)
\end{align}
In eqs~(\ref{Eq2}) -- (\ref{Eq5}), points represent particles, and lines are bonds between them such that they have an interaction. Each bond corresponds to a Mayer function in the integrals. If there are no bonds, then there are no interactions ($f_{ij}=0$). Clearly, the calculation of $B_m(T)$ involves integration over a large number of $m$-particle irreducible graphs. The number of such graphs increases exponentially with $m$, making the calculation of higher-order $B_m(T)$s cumbersome. By definition, $B_1(T)=1$ and $B_m(T)=0$ for $m>1$ for ideal gases and nonzero for imperfect gases. Thus, $B_m(T)$s act as a measure of deviation from the ideal gas behavior.

There exist theoretical and numerical evaluations of virial coefficients for many model systems~\cite{bm10}. For piece-wise constant potentials, e.g., hard-sphere, square-well, etc., an exact analytical calculation of $B_m(T)$s is possible up to a certain order. For example, Boltzmann analytically calculated $B_{3}(T)$ and $B_{4}(T)$ in spatial dimension $d=3$ for hard spheres~(HS)~\cite{bvn99}. Later, Luban and Barum extended the calculation of $B_{3}(T)$ for an arbitrary $d$~\cite{lb92}. Clisby and McCoy calculated $B_{4}(T)$ in $d$ = 4, 6, 8, 10, 12~\cite{cm04}, and Lyberg in $d$ = 5, 7, 9, 11~\cite{il05} for the same system. For attractive square-well~(SW) potential, Kihara analytically obtained $B_2(T)$ and $B_3(T)$ for arbitrary range of attraction $\lambda$ ~\cite{tk53}. Katsura calculated $B_4(T)$ for $\lambda$ equals the molecular diameter~\cite{sk59}. Later, Barker and Monaghan observed some  discrepancies in  Katsura's work and proposed a more accurate method~\cite{bm62} to evaluate $B_4(T)$. The rest of the higher-order virial coefficients, i.e., $B_m(T)$s for $m>4$, are evaluated numerically. For example, Ree and Hoover calculated $B_5(T)$ to $B_7(T)$~\cite{rh64,rh67}, van Rensburg calculated $B_8(T)$~\cite{er93}, Labik \textit{et al.} calculated $B_9(T)$~\cite{lkm05}, and Clisby and McCoy calculated $B_{10}(T)$ for HS systems for different spatial dimensions~\cite{cm06}. For SW potential, Do \textit{et al.} calculated $B_5(T)$ to $B_9(T)$ numerically~\cite{dfs16}. 

For continuous nonlinear potentials, e.g., Lennard-Jones~(LJ), repulsive inverse power, etc., an exact calculation of virial coefficients is difficult. For LJ potential [$U_{\rm LJ}(r)= (M/r^\alpha) - (N/r^\beta)$], Garrett obtained $B_2(T)$ in a closed form for $\alpha = 2\beta$~\cite{ag80}. Later, Kihara calculated $B_3(T)$ for $N,M>0$ and $\alpha>\beta>3$~\cite{tk53}. Using numerical methods, Barker \textit{et al.} calculated up to $B_5$ for LJ systems in $d=2$ and $d=3$~\cite{jb80,blp66}. Singh and Kofke calculated from $B_2(T)$ to $B_6(T)$~\cite{sk04} using the umbrella sampling method for the standard LJ system. Later, Schultz and Kofke extended the study up to $B_8(T)$~\cite{sk09}, and recently, Feng \textit{et al.} to $B_{16}(T)$~\cite{fsc04}. For repulsive inverse-power-law potential~[$U_{\rm LJ}(r)$ with $N=0$], Kihara obtained $B_2(T)$ and $B_3(T)$~\cite{tk53}, and Dixon and Hutchinson calculated $B_4(T)$ for $\alpha\geq4$ analytically~\cite{mp79}. Wheatley obtained from $B_3(T)$ to $B_7(T)$ numerically~\cite{rw13}. Later, Kofke extended the study up to $B_8(T)$~\cite{tsk11}. Also, there are evaluations of virial coefficients for Yukuwa potential and water molecules~\cite{ns09,bsk07,bss07}.

In the past, virial EoS has been used in many research areas in physics, ranging from thermodynamics~\cite{bcn92,mmvm03,NS2004} to nuclear physics~\cite{HS2006} to astrophysics~\cite{sho11}. The renewed interest in studying the virial EoS is due to advances in theory, computing power, molecular modeling and machine-learning techniques~\cite{sk2022}. For example, MacDowell \textit{et al.} investigated the virial EoS for quadrupolar Lennard-Jones diatomics for a number of quadrupole moments and elongations, considering up to fourth-order virial coefficient~\cite{mmvm04}. They determined the liquid-vapor phase transition's critical temperature and pressure, which accord well with perturbation theory predictions. Mon recently used virial expansion to study the longitudinal pressure equation for hard spheres in small cylindrical pores~\cite{mon18}. For narrow pores, the analytical results show very good agreement for a wide range of pressures with the Monte Carlo data. Shen \textit{et al.} calculated the virial EoS for a nonideal gas consisting of neutrons, protons, $\alpha$-particles, and thousands of heavy nuclei at a subnuclear density~\cite{sht10}.

It is well known that the evolution of the density, velocity, and granular temperature fields in the case of granular gases interacting via binary collisions is well represented by the macroscopic hydrodynamic equations for hard spherical fluids~\cite{NBbook}. In contrast, for dense granular flows, particles remain in touch for extended periods of time, and the likelihood of three-body or more collisions is not insignificant. Finding a coarse-grained description for such dense granular flows has been an important and challenging issue in the physics of granular materials. This kind of coarse-grained representation often necessitates a link between temperature $T$, number density $n$, and pressure $P$. For hard-sphere interaction, such a relationship has been obtained by using virial expansion~\cite{pb21,CarH}. However, a similar study of the virial EoS for a model granular interaction that can describe dense granular flow is still missing, to the best of our knowledge. In this paper, we study virial EoS for a granular potential $\Phi(r)$ as described in Sec.~\ref{sec2}. Das \textit{et al.} proposed this potential to study dense granular flows. Next, we integrate the expression for $B_m(T)$s using multidimensional Monte-Carlo integration [Sec.~\ref{sec2a}] and importance sampling [Sec.~\ref{sec2b}] methods in $d=3$. In Sec.~\ref{sec3}, we use molecular dynamics simulation for a system of particles interacting via $\Phi(r)$ to calculate pressure ($P$) as a function of density $n$ and temperature $T$ and compare with the virial equation. Finally, we summarize our results in Sec.~\ref{sec4}.
\begin{figure}
    \centering
    \includegraphics[width=0.55\textwidth]{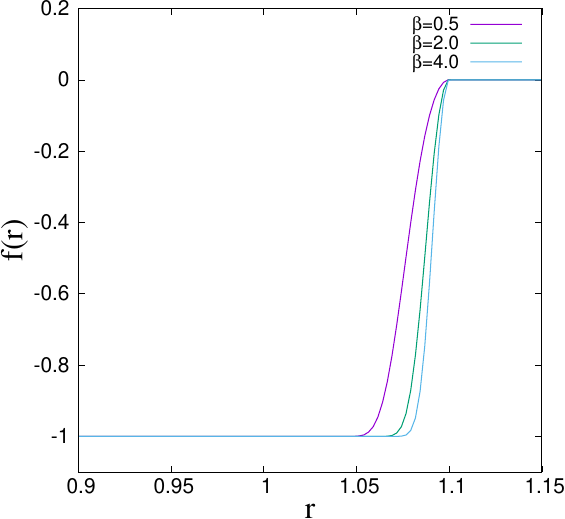}
    \caption{Plot of $f(r)$ vs. $r$ for $\Phi(r)$ for different inverse temperature $\beta$.}
    \label{fig1}
\end{figure}

\section{Numerical Estimation}\label{sec2}
We begin with a description of our model interaction potential. Two particles with position vectors $\vec r_i$ and $\vec r_j$ interact via a two-body potential of the following form:
\begin{eqnarray}
\label{pot}
    \Phi(r_{i j})= \begin{cases}\infty & ; r_{i j}<R_1, \\ V_0 \frac{\left(r_{i j}-R_2\right)^2}{\left(r_{i j}-R_1\right)^2} & ; R_1 \leqslant r_{i j}<R_2, \\ 0 & ;r_{i j} \geqslant R_2,\end{cases}
\end{eqnarray}
Here, $r_{ij}=|\boldsymbol{r}_i - \boldsymbol{r}_j|$ is the separation between the two particles, and $V_0$ is the amplitude of the potential. This potential represents a hard core of diameter $R_1$ surrounded by a thin repulsive interaction potential of diameter $R_2$ such that $R_2 - R_1 < R_1$. Eq.~(\ref{pot}) models a repulsive potential that rises steeply from 0 at the outer boundary of the shell to infinity at the hard core, mimicking a granular interaction potential~\cite{dps16}. Moreover, $\Phi(r)$ reduces to the hard sphere potential in the limit $R_2\rightarrow R_1$. In Fig.~\ref{fig1}, we plot the Mayer function $f(r)$ as a function of pair-distance $r$ for $\Phi(r)$ in eq.~(\ref{pot}) for different $\beta$. Clearly, $f(r)$ is nonzero only for $r\le R_2$.
\subsection{Direct Integration Method}\label{sec2a}
We calculate virial coefficients in eqs~(\ref{Eq3}) -- (\ref{Eq5}) by solving cluster integrals using the multidimensional Monte-Carlo approach in $d=3$. Therefore, we need to solve a $dm$-dimensional integration to evaluate $B_m(T)$. Since the interaction potential $\Phi(r)$ is radially symmetric,  we will reduce the number of integration variables by performing appropriate variable transformations. This will improve the efficiency of our numerical integrations and the convergence of our results.

To calculate $B_2(T)$, we place one particle at the origin of a spherical polar coordinate system and allow the other particle in the pair to adopt any position or orientation. The calculation for $B_2(T)$ comes down to
\begin{equation}
\label{Eq6}
B_2=-\frac{1}{2V}\iint f_{12} d^3 r_1 d^3 r_2=-\frac{1}{2}\int f_{12}(r_{12})d^3r_{12}=-2\pi\int_{0}^\infty f_{12}(r_{12})r_{12}^2 dr_{12},
\end{equation}
which is a one-dimensional integral. We set the upper limit of the integral to $r_{12}^{\rm max}=1.5$ because the integrand in the above equation is zero for $r_{12}>1.1$. We obtain $B_2(T)$ by averaging over 10,000 Monte-Carlo runs. We wish to emphasize that we also calculated $B_2(T)$ by altering $r_{12}^{\rm max}$ and found no significant variance.

The calculation of $B_{3}$ involves integration over three-particle clusters, as given by eq.~(\ref{Eq4}). To assess such integrals, we fix a particle p1 at the origin of the spherical polar coordinate system and allow the second particle p2 to move solely along the z-axis, as illustrated in Fig.~2. We enable the third particle, p3, to take any feasible positions for a given position of p2. Finally, we repeat the procedure for different positions of p2~\cite{asm18}. As a result, the eq.~(\ref{Eq4}) can be expressed as
\begin{equation}
B_3=-\frac{1}{3 V} \iiint f_{12} f_{13} f_{23} d^3 r_1 d^3 r_2 d^3 r_3=-\frac{1}{3 } \iint f_{12} f_{13} f_{23} d^3 r_{12} d^3 r_{13}.
\label{Eq7}
\end{equation}
Again, $\left|\boldsymbol{r}_{13}-\boldsymbol{r}_{12}\right|=r_{23}=\sqrt{r_{12}^2+r_{13}^2-2 r_{12} r_{13} \cos \phi}$, $d^3r_{13} = 2\pi r_{13}^2 dr_{13} \sin \phi d\phi$, and $d^3r_{12} = 4 \pi r_{12}^2 dr_{12}$. Therefore, eq.~(\ref{Eq7}) reduces to
\begin{equation}
B_3(T)\!=\!-\frac{8\pi^2}{3}\left[\int_0^{\infty}\!\!\!dr_{12}\int_0^{\infty}\!\!\!dr_{13}\int_{0}^{\pi}\!\!\!d\phi f(r_{12})
f(r_{13}) f\left(\sqrt{r_{12}^2+r_{13}^2-2 r_{12} r_{13}\cos \phi} \right)r_{13}^2 r_{12}^2 \sin\phi\right],
\end{equation}
which is a three-dimensional integral. We choose $r_{12}^{\rm max}=r_{13}^{\rm max}=2$ in our numerical simulation and obtain $B_3(T)$ by averaging over 50,000 Monte-Carlo runs.
\begin{center}
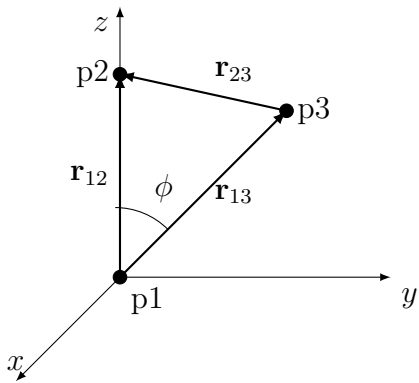

\begin{tikzpicture}[scale=1.8, >=latex]
    % Draw the coordinate axes
    \draw[->] (0,0,0) -- (2,0,0) node[below right] {$y$};
    \draw[->] (0,0,0) -- (0,2,0) node[below left] {$z$};
    \draw[->] (0,0,0) -- (0,0,2) node[above] {$x$};
    
    % Place the beads
    \filldraw (0,0,0) circle (0.05) node[below right] {p1};
    \filldraw (0,1.5,0) circle (0.05) node[left] {p2};
    \filldraw (2,2,2) circle (0.05) node[right] {p3};
    
    % Draw lines and vectors
    %\draw[dashed] (0,0,0) -- (0,1.5,0)
    \draw[->, thick] (0,0,0) -- (0,1.5,0) node[left, midway, left] {$\mathbf{r}_{12}$};
    \draw[->, thick] (2,2,2) -- (0,1.5,0) node[midway,above right] {$\mathbf{r}_{23}$};
    \draw[->, thick] (0,0,0) -- (2,2,2) node[pos=.5,right,right] {$\mathbf{r}_{13}$};
  % Draw the arc
   \draw (0,0,0) ++(45:0.5) arc[start angle=45, end angle=90, radius=0.55] node[midway, above right] {$\phi$};
\end{tikzpicture}
\captionof{figure}{A schematic diagram showing the interaction among three particles.}
\end{center}
\label{Fig2}

The computation of $B_4(T)$ is linked to the assessment of four-particle cluster integrals, as provided by eq.~(\ref{Eq5}). To compute $B_4(T)$, we utilize an extended version of the technique used to calculate $B_3(T)$. In this scenario, the four particles are arranged as follows: First, we fix a particle p1 at the spherical polar coordinate's origin $o$ and constrain another particle p2 to move solely along the $z$-axis. Then, as illustrated in Fig.~3, we position particles p3 and p4 at $i$ and $k$, respectively. As a result, the eq.~(\ref{Eq5}) can be expressed in terms of $r_{ij}$, $\theta_{ij}$, and $\phi$. Here, $r_{ij}$ represents the distance between the particles $i$ and $j$, $\theta_{ij}$ represents the angle between the vectors $\boldsymbol{r}_{oi}$ and $\boldsymbol{r}_{oj}$, and $\phi$ represents the angle between the $oij$-plane and $ojk$-plane. Again, $r_{ij}^2=r_{oi}^2+r_{oj}^2-2r_{oi}r_{oj}\cos\theta_{ij}$ with $i\neq j\neq o$, $\cos \theta_{ik}=\cos \theta_{ij} \cos \theta_{jk}+\sin \theta_{ij} \sin \theta_{jk} \cos \phi$~\cite{bm62,msa16}.
\begin{align}
B_4(T)&=-\frac{1}{8V}\iiiint \left[3f_{13}f_{14}f_{23}f_{24} + 6f_{12}f_{13}f_{14}f_{23}f_{24} + f_{12}f_{13}f_{14}f_{23}f_{24}f_{34}\right]d^3r_{1}d^3r_{2}d^3r_{3}d^3r_{4}\nonumber \\
&=-\frac{1}{8}\iiint \left[3f_{13}f_{14}f_{23}f_{24} + 6f_{12}f_{13}f_{14}f_{23}f_{24} + f_{12}f_{13}f_{14}f_{23}f_{24}f_{34}\right]d^3r_{12}d^3r_{13}d^3r_{14}.
\label{Eq9}
\end{align}

\begin{center}
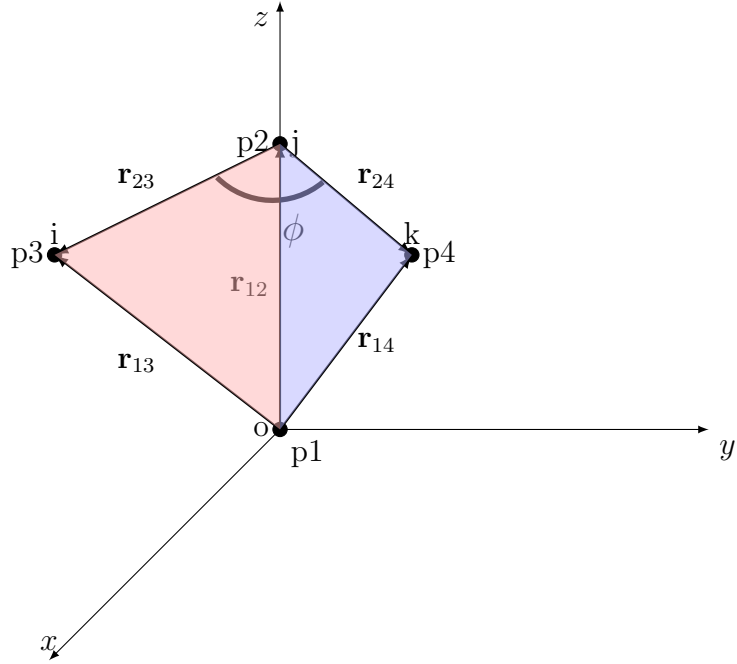

\begin{tikzpicture}[scale=1.9, >=latex]
    % Draw the coordinate axes
    \draw[->] (0,0,0) -- (3,0,0) node[below right] {$y$};
    \draw[->] (0,0,0) -- (0,3,0) node[below left] {$z$};
    \draw[->] (0,0,0) -- (0,0,4.2) node[above] {$x$};
    % Place the beads
    \filldraw (0,0,0) circle (0.05) node[below right] {p1} node[left] {o};
    \filldraw (0,2,0) circle (0.05) node[left,above,left] {p2} node[right] {j};
    \filldraw (1.5,1.8,1.5) circle (0.05) node[right] {p4} node[left,above] {k};
    \filldraw (-1,1.8,1.5) circle (0.05) node[left] {p3} node[left,above] {i};
    % Draw vector lines
    \draw[->, thick] (0,0,0) -- (0,2,0) node[left, midway, left] {$\mathbf{r}_{12}$};
    \draw[->, thick] (0,2,0) -- (1.5,1.8,1.5) node[midway,above right] {$\mathbf{r}_{24}$};
    \draw[->, thick] (0,0,0) -- (1.5,1.8,1.5) node[pos=.5,right,right] {$\mathbf{r}_{14}$};
    % Draw lines connecting p1, p3, and p4
    \draw[->, thick] (0,0,0) -- (-1,1.8,1.5) node[midway,below left] {$\mathbf{r}_{13}$};
    \draw[->, thick] (0,2,0) -- (-1,1.8,1.5) node[midway, above left] {$\mathbf{r}_{23}$};
    % Draw the arc
    \draw[line width=2pt] (0,2,0) ++(-152:0.5) arc[start angle=225, end angle=310, radius=0.55] node[midway, below right, font=\large] {$\phi$};
    % Fill the planes
    \fill[blue!30, opacity=0.5] (0,0,0) -- (1.5,1.8,1.5) -- (0,2,0) -- cycle;
    \fill[red!30, opacity=0.5] (0,0,0) -- (-1,1.8,1.5) -- (0,2,0) -- cycle;
\end{tikzpicture}
\captionof{figure}{A schematic diagram showing a four-body cluster.}
\end{center}
\label{Fig3}
Further, the volume elements in eq.~(\ref{Eq9}) reduce to $d^3 r_{14}=r_{14}^2\sin\theta_{24}dr_{14}d\theta_{24}d\phi$, $d^3r_{13}=2\pi r_{13}^2\sin\theta_{23}d r_{13}d\theta_{23}$, and $d^3r_{12} =4\pi r_{12}^2dr_{12}$. Hence, the twelve-dimensional integration in eq.~(\ref{Eq5}) reduces to a six-dimensional integration in eq.~(\ref{Eq9}). Now the 1st term on RHS of eq.~(\ref{Eq9}) can be written as
\begin{align}
D_1(T)=&-\frac{3}{8} \iiint f(r_{13}) f(r_{14}) f(r_{23}) f(r_{24}) d^3 r_{12} d^3 r_{13} d^3 r_{14} \nonumber \\
=&-6\pi^3\int_0^{\infty} \int_0^{\infty} \int_0^{\infty} \int_{-1}^1 \int_{-1}^1 r_{12}^2 f\left(r_{13}\right)r_{14}^2 f\left(r_{14}\right) r_{13}^2 f\left(\sqrt{r_{12}^2+r_{13}^2-2 r_{12} r_{13} \eta_{23}}\right)\nonumber \\
&\times f\left(\sqrt{r_{12}^2+r_{14}^2-2 r_{12} r_{14}\eta_{24}}\right) d r_{12} d r_{13}d r_{14} d \eta_{23} d \eta_{24}.
\label{Eq10}
\end{align}
The 2nd term on RHS of eq.~(\ref{Eq9}) becomes
\begin{align}
D_2(T)=&-\frac{3}{4} \iiint f(r_{12})f(r_{13}) f(r_{14}) f(r_{23}) f(r_{24})  d^3 r_{12} d^3 r_{13} d^3 r_{14}\nonumber \\
=&-12 \pi^3 \int_0^{\infty} \int_0^{\infty} \int_0^{\infty} \int_{-1}^1\int_{-1}^1 r_{12}^2 f\left(r_{12}\right) r_{13}^2 f\left(r_{13}\right) r_{14}^2 f\left(r_{14}\right) f\left(\sqrt{r_{12}^2+r_{13}^2-2 r_{12} r_{13} \eta_{23}}\right)\nonumber \\
&\times f\left(\sqrt{r_{12}^2+r_{14}^2-2 r_{12} r_{14}\eta_{24}}\right) d r_{12} d r_{13}d r_{14} d \eta_{23} d \eta_{24}.
\label{Eq11}
\end{align}
And the last term on RHS of eq.~(\ref{Eq9}) becomes
\begin{align}
D_3(T)=&-\frac{1}{8} \iiint f(r_{12}) f(r_{13}) f(r_{14}) f(r_{23}) f(r_{24}) f(r_{34}) d^3 r_{12} d^3 r_{13} d^3 r_{14}\nonumber \\
= & -\pi^2 \int_0^{\infty} \int_0^{\infty} \int_0^{\infty} \int_{-1}^1\int_{-1}^1\int_0^{2\pi} r_{12}^2 f\left(r_{12}\right) r_{13}^2 f\left(r_{13}\right) r_{14}^2 f\left(r_{14}\right) f\left(\sqrt{r_{12}^2+r_{13}^2-2 r_{12} r_{13} \eta_{23}}\right)\nonumber \\
&\times f\left(\sqrt{r_{13}^2+r_{14}^2-2 r_{13} r_{14}\left\{\eta_{23} \eta_{24}+\sqrt{\left(1-\eta_{23}^2\right)} \sqrt{\left(1-\eta_{24}^2\right)} \cos \phi\right\}}\right)\nonumber \\
&\times f\left(\sqrt{r_{12}^2+r_{14}^2-2 r_{12} r_{14}\eta_{24}}\right)dr_{12}dr_{13}dr_{14}d\eta_{23}d\eta_{24}d\phi.
\label{Eq12}
\end{align}
Here, we have used $\eta_{23}=\cos\theta_{23}$ [i.e., $\eta_{23}\in (-1,+1)$],  $\eta_{24}=\cos\theta_{24}$ [i.e., $\eta_{23}\in (-1,+1)$], and 
\begin{align}
   \eta_{34}=\cos \theta_{34} =\cos \theta_{23} \cos \theta_{24}+\sin \theta_{23} \sin \theta_{24} \cos \phi =\eta_{23} \eta_{24}+\sqrt{(1-\eta_{23}^2)(1-\eta_{24}^2)} \cos\phi.
\end{align}
We obtain $B_4(T)$ by solving eqs.~(\ref{Eq10}) -- (\ref{Eq12}) using Monte-Carlo procedure. We use $r_{12}^{\rm max}=r_{13}^{\rm max}=r_{14}^{\rm max}=2.0$ and the result is averaged over 50,000 independent runs. Table~\ref{10table:1} shows the values of $B_2(T)$, $B_3(T)$ and $B_4(T)$ at different temperatures.
\begin{table}
    \centering
    \renewcommand{\arraystretch}{1.6} % Increase the row spacing
    \caption{$B_m(T)$s obtained from direct integration.}
    \label{10table:1}
    \begin{tabular}{| c | c | c | c |}
        \hline  
        $T$ & $B_2$ & $B_3$ & $B_4$ \\ [1ex] \hline
        1 & 2.654310 & 4.402244 & 5.359088 \\ \hline
        2 & 2.615108 & 4.272537 & 5.122250 \\ \hline
        3 & 2.589089 & 4.187514 & 4.968951 \\ \hline
        4 & 2.569371 & 4.123662 & 4.854866 \\ \hline
        5 & 2.553433 & 4.072416 & 4.763967 \\ \hline
        6 & 2.540037 & 4.029603 & 4.688484 \\ \hline
        7 & 2.528479 & 3.992851 & 4.624024 \\ \hline
        8 & 2.518314 & 3.960674 & 4.567848 \\ \hline
        9 & 2.509243 & 3.932078 & 4.518129 \\ \hline
        10 & 2.501057 & 3.906363 & 4.473585 \\ \hline
    \end{tabular}
\end{table}

\subsection{Importance Sampling Method}\label{sec2b}
We calculate $B_2(T)$, $B_3(T)$, and $B_4(T)$ directly from eqs.~(\ref{Eq3}) -- (\ref{Eq5}) utilizing the \textit{importance sampling}~(IS) procedure~\cite{fs01,gtc07}. In IS, we sample the configuration space for a $m$-particle cluster according to some distribution  $\pi\left(\boldsymbol{r}^m ; T\right)$ (may be unnormalized) with non-zero integrands. IS advices that we use $\pi\left(\boldsymbol{r}^m ; T\right)=\left|\gamma\left(\boldsymbol{r}^m ; T\right)\right|$ for better convergence of the integrations, where $\gamma\left(\boldsymbol{r}^m ; T\right)$ is the integrand in the expression of $B_m(T)$. The \textit{Umbrella Sampling}~(US)~\cite{fs01} is then used to calculate a generic cluster integral $\Gamma(T)$ corresponding to the $m^{\rm th}$ virial coefficient.  The US technique necessitates the use of a reference system for which $\Gamma_0$ (say) is the corresponding virial coefficient, after which we can write~\cite{bsk07,sk04,sbc13}
\begin{equation}
\label{Eq14}
\Gamma(T)=\Gamma_0 \frac{\left\langle\gamma/\pi\right\rangle_\pi}{\left\langle\gamma_0 / \pi\right\rangle_\pi},
\end{equation}
where $\gamma_0\left(\boldsymbol{r}^m ; T\right)$ is the integrand in the expression of $B_m(T)$ for the reference system. For instance, if $\Gamma$ is $B_3$, then $\gamma = f_{12} f_{13} f_{23}$. The angular brackets represent an average over the configuration space $\pi\left(\boldsymbol{r}^m ; T\right)$. Since we choose $\pi=\left|\gamma\left(\mathbf{r}^m ; T\right)\right|$, then eq.~(\ref{Eq14}) reduces to the following form:
\begin{equation}
\label{Eq15}
\Gamma(T)=\Gamma_0 \frac{\left\langle {sign}\right\rangle_\pi}{\left\langle\gamma_0 / \pi\right\rangle_\pi},
\end{equation}
where ${sign}$ can only take values $+1$ or $-1$. 

Next, we select a reference system that has a high phase space overlap with the target system. As our reference system, we use the square well~(SW) interaction potential, given by
\begin{equation}
\label{Eq16}
U\left(r_{i j}\right)= \begin{cases}\infty &; 0<r_{i j}<R_1 \\ -\varepsilon &; R_1 \leq r_{i j}\leq R_2, \\ 0 &; R_2 < r_{ij}.\end{cases}
\end{equation}
Here, $R_1$ is the diameter of the particles and $\epsilon$ is the strength of SW potential. The theoretical values of the first few virial coefficients for SW potential in $d=3$ are~\cite{sm91}
\begin{align}
B_2/b^{~}=&1.000-0.331 f, \\
B_3/b^2=&0.6250 - 0.3762f + 0.9906\times 10^{-1}f^2 - 0.6946\times 10^{-2}f^3, \\
B_4/b^3=&0.2869 - 0.2207f + 0.1209f^2 - 0.4131\times 10^{-1}f^3 + 0.6438\times 10^{-2}f^4 \nonumber \\ &-0.3090\times 10^{-3}f^5-0.3023\times 10^{-5}f^6,
\end{align}
where $\textit{b}=2\pi/3$ with $R_1=1$, $R_2=1.1$ and $f=\exp{(\beta\epsilon)} - 1$.

To build the configuration space according to the distribution $\pi\left(\boldsymbol{r}^m; T\right)$, we generate a random point using a uniform random number generator in $3m$-dimensional space. We select the first point~(old) at the origin of the $3m$-dimensional space. All other points~(new) are generated by keeping three coordinates (coordinates of one fixed particle) at the origin and the rest of the $3(m-1)$ coordinates (rest of the ($m-1$) particles of the cluster) by randomly perturbing between $-R_2$ and $+R_2$ with respect to the existing positions. The acceptance probability $p$ is then calculated using the Metropolis Monte Carlo algorithm~\cite{jsl04} as $p = \min\left(1, \pi^{\rm new} / \pi^{\rm  old}\right)$. Replace the old point with the new point if it is accepted. We continue the preceding stages until we have generated a sufficient number of points in the configuration space based on the desired distribution $\pi\left(\boldsymbol{r}^m; T\right)$.

\begin{table}
    \centering
    \renewcommand{\arraystretch}{1.2} % Increase the row spacing
    \caption{$B_m(T)$s using importance sampling for different $\epsilon$.}
    \label{10table:2}
    \begin{tabular}{|c|*{9}{p{1.37cm}|}}
        \hline
        & \multicolumn{3}{c|}{\centering $\epsilon=0.1$ } & \multicolumn{3}{c|}{\centering $\epsilon=0.01$ } & \multicolumn{3}{c|}{\centering $\epsilon=0.001$ } \\
        \hline
        \multicolumn{1}{|c|}{$T$} & \multicolumn{1}{c|}{$B_2$} & \multicolumn{1}{c|}{$B_3$} & \multicolumn{1}{c|}{$B_4$} & \multicolumn{1}{c|}{$B_2$} & \multicolumn{1}{c|}{$B_3$} & \multicolumn{1}{c|}{$B_4$} & \multicolumn{1}{c|}{$B_2$} & \multicolumn{1}{c|}{$B_3$} & \multicolumn{1}{c|}{$B_4$} \\
        \hline
        1 & 2.656056 & 4.413944 & 5.214216 & 2.655434 & 4.408325 & 5.260785 & 2.655377 & 4.407819 & 5.265290 \\
        \hline
        2 & 2.617553 & 4.284999 & 5.075579 & 2.617144 & 4.282513 & 5.093469 & 2.617104 & 4.282263 & 5.095186 \\
        \hline
        3 & 2.592008 & 4.193893 & 4.984408 & 2.591528 & 4.192988 & 4.997902 & 2.591482 & 4.192901 & 4.999225 \\
        \hline
        4 & 2.572503 & 4.143922 & 4.899601 & 2.572103 & 4.141260 & 4.900358 & 2.572064 & 4.141000 & 4.900460 \\
        \hline
        5 & 2.556372 & 4.089478 & 4.856408 & 2.556112 & 4.087747 & 4.831824 & 2.556086 & 4.087577 & 4.829466 \\
        \hline
        6 & 2.543169 & 4.047274 & 4.758576 & 2.542882 & 4.045256 & 4.746167 & 2.542853 & 4.045058 & 4.744959 \\
        \hline
        7 & 2.531828 & 4.017103 & 4.678109 & 2.531563 & 4.014324 & 4.663105 & 2.531537 & 4.014050 & 4.661648 \\
        \hline
    \end{tabular}
\end{table}

\begin{figure}
\includegraphics[width=0.99\textwidth]{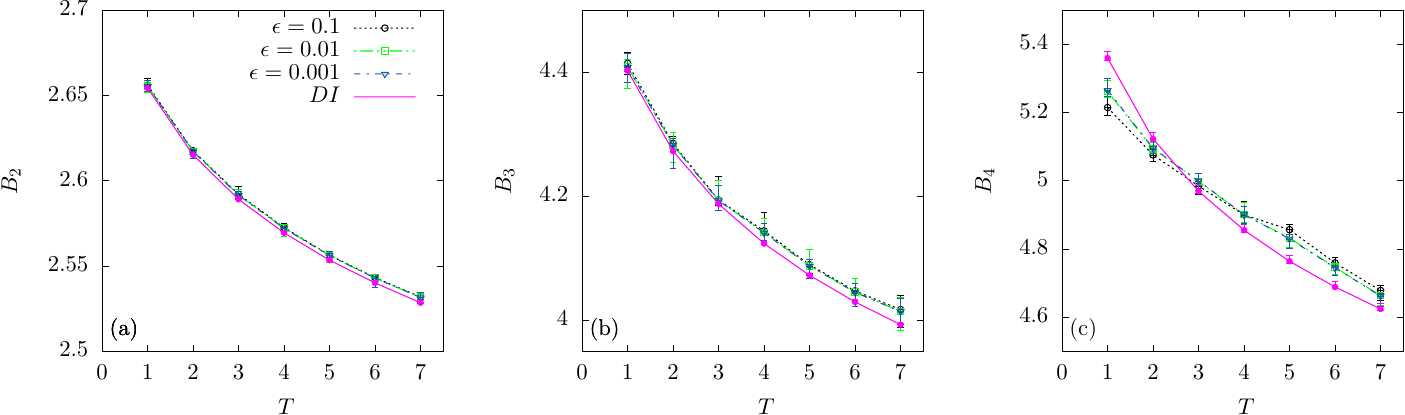}
\caption{Plot of virial coefficients as a function of temperature $T$. (a) Plot of $B_2$ vs. $T$. The magenta solid line represents the result of the direct integration~(DI) method. As indicated, the different line points represent data gathered using the importance sampling~(IS) approach for different $\epsilon$ values. (b) Plot of $B_3$ vs. $T$, and (c) Plot of $B_4$ vs. $T$. The symbols used in (b) and (c) have the same meaning as the ones in (a).}
\label{Fig4}
\end{figure}

Table~\ref{10table:2} provides the numerical values of $B_2(T)$, $B_3(T)$, and $B_4(T)$ at different temperatures for three different values of $\epsilon$ obtained by importance sampling method. We notice no substantial variation in $B_i(T)$ as $\epsilon$ varies for a given $T$. The $B_i(T)$ values obtained from the DI and IS approaches are presented in Fig.~\ref{Fig4}. We notice that all the virial coefficients obtained by IS and DI approaches are extremely close. Further, we want to emphasize that in all the above calculations, $T$ can be chosen freely.

Let us now discuss the advantages and disadvantages of DI and IS. In DI, we reduce the multidimensional integrations associated with the expression of virial coefficients for better convergence of numerical results. Such reduction often requires complex transformations of coordinates as we go for higher-order virial coefficients. No such reduction is required in the case of IS. In IS, we need a reference system to calculate $B_i$s of a target system. We choose a reference system in such a way that there is a significant overlap in the configuration spaces of both systems. In this case, the estimation of $B_i$s for the target system requires the value $B_i$ for the reference system in advance, which may be approximated earlier. The DI method does not require any reference system, and thus it is free from all kind of pre-approximations. Further, we can generalize the calculation of $B_i$s to any arbitrary order $i$ in the case of IS as long as we know the corresponding $B_i$ of the reference system. No such straight-forward generalization for higher-order virial coefficients is possible for DI.

\section{Comparison with Molecular Dynamics Results}\label{sec3}
We obtain $P$ for a system of particles interacting via $\Phi(r)$ provided by eq.~(\ref{pot}) using molecular dynamics~(MD) simulation. In this case, the normal force acting on the $i^{\rm th}$ particle due to the  $j^{\rm th}$ particle is given by
\begin{eqnarray}
\label{fn}
\vec F_{ij}(r) = -\vec \nabla_j \Phi(r),
\end{eqnarray}
where $\vec \nabla_j$ is the gradient with respect to $\vec r_j$. We employ the following units for various relevant quantities: lengths are expressed in units of $R_1$, energy in $u=V_0/15$, temperature in $u/k_{\rm B}$, and time in $t_0=\sqrt{m_0R_1^2/V_0}$, where $m_0$ is the mass of a particle. We choose $V_0=15$, $R_1=1$, and $R_2=1.1R_1$ since granular interactions are inherently short-ranged. We perform molecular dynamics simulations with $N = 100,000$ particles in three dimensions. We modify the linear size of the system $L$ to achieve the desired number density, $n=N/L^3$. Given that the effective diameter of a particle is $R_2=1.1$, the jamming number density for our model is $n_{\rm J}=0.918$. We have carried out numerical simulations up to $n=0.80$. In all directions, periodic boundary conditions are used. We use the velocity-Verlet algorithm to update positions and velocities with a time step of $\Delta t = 0.0005$. The temperature $T$ of the system is given by the relation
\begin{equation}
    \frac{3}{2}NT = \sum_{i=1}^{N}\frac{1}{2}m_0\vec v_i^{\,2}.
    \label{Eq22}
\end{equation}
In earlier work, Das \textit{et al.} used this interaction potential to study freely evolving granular gases~\cite{dps16}, heated granular systems in the low and high-density limits~\cite{dps17}, and the dynamics of an intruder in granular media~\cite{dps20}.

\begin{figure}
\includegraphics[width=.99\textwidth]{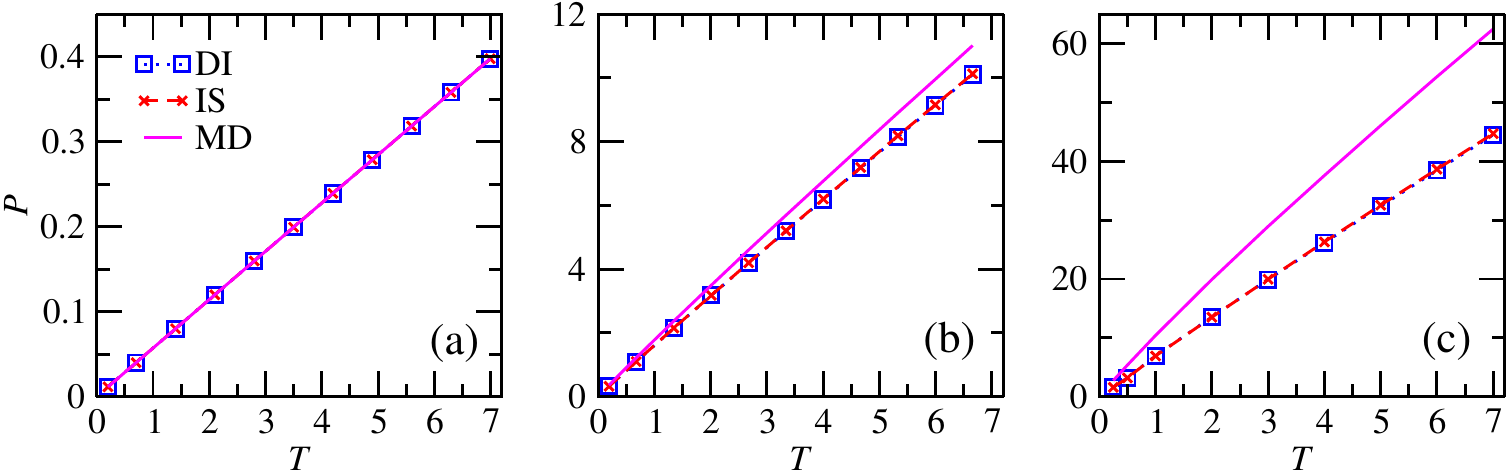}
\caption{Pressure $P$ vs. Temperature $T$ plotted for different densities (a) $n=0.05$, (b) $n=0.45$, and (c) $n=0.80$. The blue line-points represent a plot of eq.~(\ref{Eq1}) with $B_m(T)$s obtained from the direct integration~(DI) method, the red line-points represent plot of eq.~(\ref{Eq1}) with $B_m(T)$s obtained from the important sampling~(IS) method, and the solid magenta lines represent data from the molecular dynamics (MD).}
\label{Fig5}
\end{figure}

At $t=0$, we randomly inject particles into the simulation box so that the cores of no two particles overlap. We assign random velocities to the particles such that the centre of mass velocity of the system is equal to zero. First, we allow the system to equilibrate until $t=50$, which relaxes the velocity distribution to the Maxwell-Boltzmann velocity distribution. Next, we compute the elements of the stress tensor $\sigma^{\alpha\beta}$ within the volume $L^3$ as the sum of the contact stress and kinetic stress terms over all particles $i$ in the system,
\begin{equation}
    \label{Eq23}
    \sigma^{\alpha\beta} = \frac{1}{L^3}\sum_{i}\left[\sum_{j\ne i} \frac{r_{ij}^\alpha F_{ij}^\beta}{2} + mv_i^\alpha v_j^\beta\right],
\end{equation}
where $\alpha, \beta = x, y, z$. Finally, the pressure $P$ is given by
\begin{equation}
    \label{Eq24}
    P = \frac{1}{3}\left[\sigma^{xx} + \sigma^{yy} + \sigma^{zz}\right].
\end{equation}
Figure~\ref{Fig5} shows the plot of $P$ vs. $T$ for different densities obtained from the MD simulations and the virial EoS~[eq.~(\ref{Eq1})]. Clearly, when $T$ and $n$ increase, so does $P$. The virial EoS with the first four virial coefficients is sufficient to describe the $P$-$T$ graph obtained from MD simulations at low densities (cf. Fig.~\ref{Fig5}(a)). As shown in Fig.~\ref{Fig5}(b), numerical data from MD simulations are likewise in good agreement with the virial EoS with the first four virial coefficients for a intermediate $n=0.45$ only at low-$T$ domain. At high-$T$, we observe a significant deviation. For higher densities, e.g., $n=0.80$ which is close to $n_{\rm J}$, we see a considerable deviation at all temeperatures, as illustrated in Fig.~\ref{Fig5}(c). Because many-body interactions contribute significantly to $\sigma^{\alpha\beta}$ through the contact stress, making the higher-order terms $B_m(T)n^m$ with $m\ge 5$ in viral EoS crucial.

\begin{figure}
\includegraphics[width=0.99\textwidth]{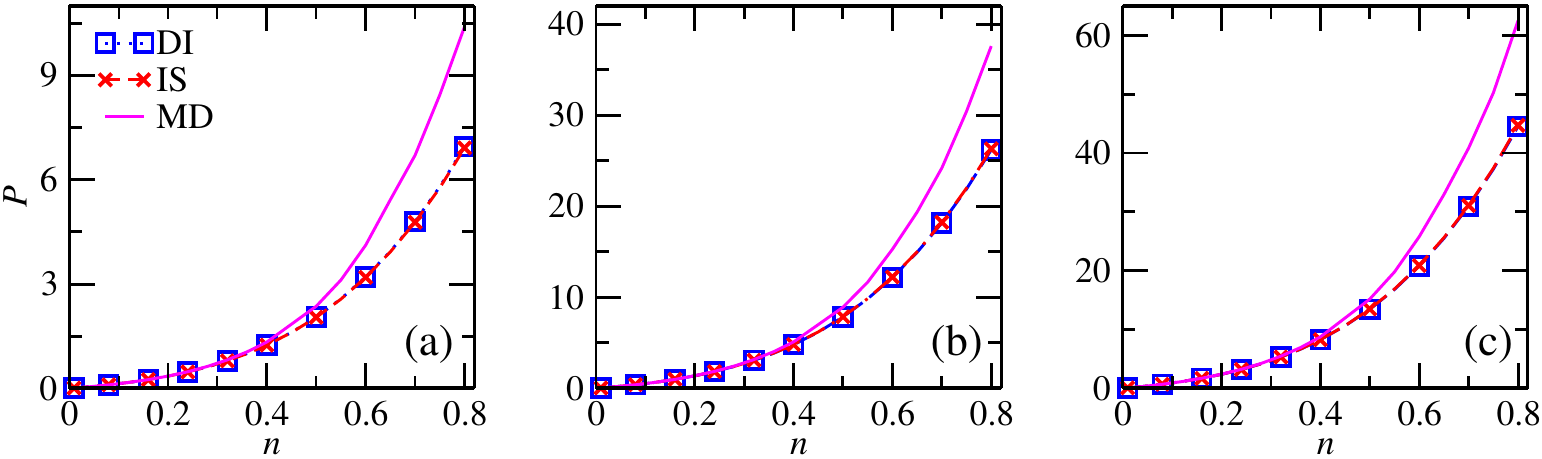}
\caption{Plot of $P$ vs. $n$ for different temperatures (a) $T=1$, (b) $T=4$, and (c) $T=7$. Symbols used here have the same meaning as the ones in Fig.~\ref{Fig5}.}
\label{Fig6}
\end{figure}

\begin{figure}
\centering
\includegraphics[width=.99\textwidth]{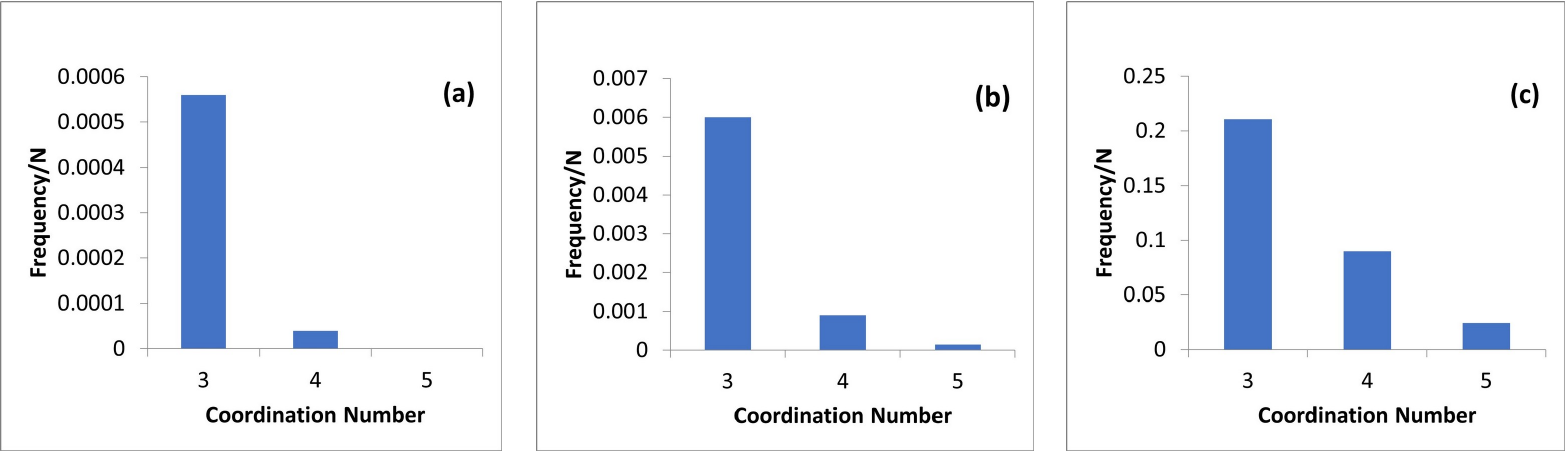}
\caption{Plot of the distribution of coordination numbers normalized by the total number of particles $N$ for different densities (a) $n=0.25$, (b) $n=0.45$, and (c) $n=0.80$ at $T=7$.}
\label{FigN}
\end{figure}

Next, we calculate the pressure $P$ at a fixed temperature $T$ as a function of density $n$. To fix $T$, we rescale the velocity of all particles so that their mean kinetic energy becomes $3T/2$ and let the system equilibrate further for time $t_{\rm eq}=10$. Following that, we calculate $T$ using eq.~(\ref{Eq22}). This process is repeated until we reach the target $T$. Then we use eqs.~(\ref{Eq23}) and (\ref{Eq24}) to calculate $P$. Figure~\ref{Fig6} shows the plot of $P$ vs. $n$ for different temperatures obtained from the MD simulations and the virial EoS~[eq.~(\ref{Eq1})]. Clearly, for all temperatures, the virial EoS with the first four virial coefficients is sufficient to describe the $P$-$n$ graph produced from MD simulations at low densities~($n<0.25$). For larger densities, MD data shows significant variation in all scenarios, highlighting the need for higher-order terms in the virial EoS. To justify our claim, we have calculated the coordination number of a particle at $T = 7$ for different densities. The distribution of coordination numbers for three densities, (a) $n = 0.25$, (b) $n = 0.45$, and (c) $n = 0.80$ is shown in Figure~\ref{FigN}. As illustrated in Fig.~\ref{FigN}(b), it is evident that a small fraction of particles for $n=0.45$ have a coordination number greater than 4, which increases with increasing the density $n$ as shown in Fig.~\ref{FigN}(c). This indicates the necessity for $B_5(T)$ and higher-order terms in the virial EoS. However, the fraction of particles having coordination number 5 is zero for $n = 0.25$, as shown in Fig.~\ref{FigN}(a). This holds true for all densities $n\le0.30$.

\section{Summary and Outlook}\label{sec4}
Let us conclude this paper with a summary and discussion. We have provided a comprehensive study of the virial equation of state~(EoS) for a model granular system. We calculated the first four virial coefficients for an interaction potential $\Phi(r)$ describing a granular interaction. We used two approaches: (a) direct integration and (b) importance sampling. We observed that the values of $B_2(T)$, $B_3(T)$, and $B_4(T)$ determined by these methods are extremely close.

Next, we perform a molecular dynamics simulation for a system of particles interacting via the same interaction potential. For different temperatures and densities, we calculate $P$ using the virial stress tensor. For densities $n\le 0.3$, the virial EoS with the first four virial coefficients is sufficiently close to the data for $P$ obtained via MD simulations at all temperatures upto $T=7$. MD data indicates significant deviation for dense systems, demonstrating the necessity for higher-order components in the virial EoS.

In our future endeavors, we aspire to delve deeper into the exploration of the equation of state for systems comprised of asymmetric particles. The dynamics of asymmetric particles hold a pivotal place within the realm of physics, extending their influence across various disciplines. One of the prominent arenas where the significance of asymmetric particle dynamics comes to the fore is polymer physics. Polymer chains inherently possess asymmetry, and their behavior, whether in solution or within bulk materials, is intricately linked to the intricate dynamics of individual polymer segments. Furthermore, the ramifications of asymmetric particle dynamics reverberate in the domains of soft matter and complex fluids, including gels, glasses, and liquid crystals. These materials consist of constituents that are themselves inherently asymmetric, and their behavior and properties are substantially influenced by the dynamics of these particles. The impact of this influence resonates through the field, shaping our understanding of a wide array of physical processes, materials science, and diverse technological applications.

\ \\
\noindent{\bf Acknowledgments:} SH acknowledges financial support from IISER Mohali through a Junior Research Fellowship. PD acknowledges financial support from SERB, India through a start-up research grant (SRG/2022/000105).

\ \\
\noindent{\bf Data Availability:} All of the data utilized in this manuscript will be made accessible upon reasonable request.

\ \\
\noindent{\bf Author contribution statement:} PD proposed the problem and develop the program for molecular dynamics simulation. SH performed the all the analytical and numerical work with the help of PD. The paper was written by PD.

\end{document}